\documentclass[sigconf,natbib=false, 9.8pt]{acmart}

\usepackage{listings}
\usepackage{siunitx}
\usepackage{makecell}
\AtBeginDocument{%
  }
    
\setcopyright{none}                
\renewcommand\footnotetextcopyrightpermission[1]{} 

\acmISBN{978-1-4503-XXXX-X/2018/06}

\RequirePackage[
  datamodel=acmdatamodel,
  style=acmnumeric,
  ]{biblatex}

\begin{document}


\title{Splats under Pressure: Exploring Performance–Energy Trade-offs in Real-Time 3D Gaussian Splatting under Constrained GPU Budgets}


\author{Muhammad Fahim Tajwar}
 \affiliation{%
   \institution{National University of Singapore}
   \country{Singapore}
 }
 \email{fahim.tajwar@u.nus.edu}

 \author{Arthur Wuhrlin}
 \affiliation{%
   \institution{National University of Singapore}
   \country{Singapore}
   \institution{Telecom Paris}
 }
 \email{arthur.wuhrlin@u.nus.edu}

 \author{Anand Bhojan}
 \affiliation{%
  \institution{National University of Singapore}
  \country{Singapore}
 }
 \email{dcsab@nus.edu.sg}


\begin{abstract}

We investigate the feasibility of real-time 3D Gaussian Splatting (3DGS) rasterization on edge clients with varying Gaussian splat counts and GPU computational budgets. Instead of  evaluating multiple physical devices, we adopt an emulation-based approach that approximates different GPU capability tiers on a single high-end GPU. By systematically underclocking the GPU core frequency and applying power caps, we emulate a controlled range of floating-point performance levels that approximate different GPU capability tiers. At each point in this range, we measure frame rate, runtime behavior, and power consumption across scenes of varying complexity, pipelines, and optimizations, enabling analysis of power–performance relationships such as FPS–power curves, energy per frame, and performance per watt. This method allows us to approximate the performance envelope of a diverse class of GPUs, from embedded and mobile-class devices to high-end consumer-grade systems.

Our objective is to explore the practical lower bounds of client-side 3DGS rasterization and assess its potential for deployment in energy-constrained environments, including standalone headsets and thin clients. Through this analysis, we provide early insights into the performance–energy trade-offs that govern the viability of edge-deployed 3DGS systems.
\end{abstract}




\maketitle
\pagestyle{plain}
\section{Introduction}
3D Gaussian Splatting (3DGS) \cite{kerbl3Dgaussians} has recently emerged as a compelling real-time rendering technique for photorealistic scene reconstruction and novel view synthesis. Unlike traditional mesh-based representations or neural radiance fields, 3DGS achieves high-quality rendering at interactive frame rates by directly rasterizing point-based primitives (3-dimensional Gaussian point clouds) on the GPU. While this performance has made 3DGS attractive for desktop-class systems, its deployment on power- and compute-constrained devices remains underexplored.

A key open question is whether the real-time rasterization component of 3DGS, which is typically offloaded to a high-end GPU, can be feasibly run on lower end consumer edge devices. This is particularly relevant for distributed or hybrid rendering pipelines, where moving some computational load from the cloud to the client can reduce motion-to-photon delay and server-side cost. However, the rasterization phase in 3DGS is computationally intensive, involving millions of Gaussian primitives and extensive GPU throughput, raising concerns about its scalability on thinner clients.

In this work, we explore the practical feasibility of client-side 3DGS rasterization under constrained GPU conditions. Rather than targeting specific devices, we simulate several GPU configurations
by controlling the single-precision floating-point performance of a desktop GPU through systematic underclocking and power limiting. We measure key performance indicators, such as frame rate and runtime behavior, across varying scene complexities and pipeline configurations. This allows us to approximate the lower performance bounds at which 3DGS rasterization remains viable, and to characterize the trade-offs involved in pushing the technique toward more lightweight deployments.

\noindent\rule{\linewidth}{0.4pt}
\noindent\textbf{Our contributions are as follows:}
\begin{itemize}

\item We present a GPU capability emulation methodology that approximates multiple consumer GPU tiers on a single high-end GPU through controlled power and clock throttling.

\item We provide the first empirical characterization of real-time 3D Gaussian Splatting performance under constrained GPU compute budgets across multiple levels of scene complexity.

\item We analyze performance--energy trade-offs using metrics such as energy per frame and performance per watt, offering insights into the feasibility of edge-deployed 3DGS systems.
\end{itemize}

Our analysis is situated in the context of prior work on neural rendering, hybrid systems, and edge-based graphics pipelines, which we review in the following section.

\section{Related Work}

\subsection{Novel View Synthesis with 3D Gaussian Splatting}

Kerbl et al.\,2023~\cite{kerbl3Dgaussians} introduced \emph{3D Gaussian Splatting} (3DGS), a point-based neural graphics primitive that stores a scene as a set of millions of \emph{anisotropic Gaussians}.  
Each Gaussian carries position, full covariance (shape and orientation), an opacity term, and view-dependent colour represented with low-order spherical harmonics.  
Rendering is performed by a \emph{visibility-aware splatting rasterizer}:  
(1) Gaussians are binned into $16{\times}16$ screen tiles,  
(2) depth-sorted per tile, then  
(3) alpha-blended front-to-back, with early-cutoff once accumulated alpha saturates.  
All steps run entirely on the GPU and expose efficient forward \emph{and} backward passes, enabling end-to-end optimisation of both geometry and appearance.  
Compared with NeRF-style \cite{mildenhall2020nerfrepresentingscenesneural} ray marching, 3DGS achieves two orders of magnitude faster training and 30–60 FPS real-time rendering at 1080 p while matching or exceeding the visual quality of state-of-the-art neural radiance fields.\cite{kerbl3Dgaussians} 
However, 3DGS also requires much more parameters, and is still not fast and light enough to be used on standalone VR headsets.
Subsequent open-source implementation ~\texttt{gsplat} \cite{ye2025gsplat} provides better training speed and memory footprint, and is a good baseline for follow-up research, since it gathers many techniques in a single package.

\subsection{Level-of-Detail and Streaming Extensions}

Although 3DGS excels on small–medium scenes, outright storing and drawing every Gaussian is prohibitive for city-scale or network-streamed content.  
Several works therefore embed 3DGS in \emph{hierarchical, layered, or streaming} frameworks.

\paragraph{Hierarchical 3DGS}
Kerbl et al.\,2024 extend their original system with a \emph{Gaussian octree} that merges distant splats into coarser parents and refines them on-line when the camera approaches \cite{hierarchicalgaussians24}.  
This yields kilometre-scale scenes that fit into a single‐GPU budget while retaining fine detail nearby.

\paragraph{Layered / Progressive LOD}
\emph{LapisGS}~\cite{shi2024lapisgs} organizes splats into a base layer plus additive enhancement layers, enabling adaptive streaming over variable-bandwidth links.  



\subsection{Gap: Client-Side 3DGS under Power and Bandwidth Constraints}

Most prior 3DGS studies assume \emph{high-end desktop-class} GPUs at the client\cite{kerbl3Dgaussians}\cite{hierarchicalgaussians24}\cite{shi2024lapisgs}.  
No work quantifies how frame-rate scale when hardware is throttled to mid or low-range desktop GPUs.  
Our study fills this gap.  
We emulate four consumer-GPU tiers, from RTX 3050 up to RTX 4090, on a single 4090 by jointly limiting power, core clock, and memory clock until sustained FP32 throughput and bandwidth match each reference tier.  
We then measure 3DGS frame-rate and resource usage across multiple LOD settings on the common \textit{Garden} scene\cite{barron2022mipnerf360}.  
The results provide the first empirical curve relating 3DGS viability to available TFLOPS and bandwidth, informing future ports to Vulkan/Metal on mobile XR SoCs and guiding LOD/streaming system design.







\section{System Overview}

\subsection{Software Backbone}
Our prototype is implemented on top of the open-source \texttt{gsplat} Python library by Ye et al.~\cite{ye2025gsplat}, which adds a highly optimised CUDA rasterizer, mixed-precision training, and out-of-core data handling to the original 3DGS codebase.  All results in this paper therefore inherit \texttt{gsplat}’s faster convergence (1.5–2 × speed-up) and lower peak memory footprint compared with Kerbl et al.’s reference implementation.

\subsection{Layered Level-of-Detail Optimisation}
We adopt the multi-resolution (“layer”) strategy of LapisGS~\cite{shi2024lapisgs} to train a hierarchy of Gaussian splat sets that act as discrete levels of detail (LoD).  
Let $n$ be the number of layers, \(\mathcal{L}_0,\ldots,\mathcal{L}_{n-1}\).  
Training images \(\mathcal{I}\) are progressively down-sampled by powers of two to obtain \(\mathcal{I}_{n-1}\) (coarsest) up to \(\mathcal{I}_0\) (full resolution).  
We then optimise layers \emph{coarse to fine}:

\begin{enumerate}
    \item Initialise and optimise \(\mathcal{L}_{n-1}\) against \(\mathcal{I}_{n-1}\).
    \item Freeze previously trained layers; spawn new Gaussians where residual error is high.
    \item Optimise \(\mathcal{L}_{k}\) against \(\mathcal{I}_{k}\) while keeping \(\mathcal{L}_{<k}\) fixed, for \(k=n-2,\ldots,0\).
\end{enumerate}


With this method, one can control the number of gaussians by lower or increasing the number of used layers. 

\subsection{Looped Dynamic Objects via 4D Gaussian Splatting}
Static scenery alone does not capture the dynamics required in XR.  
We therefore incorporate the 4D Gaussian Splatting formulation of Wu et al.~\cite{wu20244d}.  
Each dynamic asset is represented by

\begin{itemize}
    \item a \textbf{canonical} Gaussian set \(\mathcal{G}_{\text{base}}\) (time-invariant); and
    \item a small MLP \(f_\theta(t)\) that predicts a \emph{delta} for every Gaussian parameter as a function of time \(t\in[0,1]\).
\end{itemize}

Because we target looped animations (e.g.\ swaying foliage, rotating signage), the MLP is trained on a single period and queried each frame with phase-normalized time.  
At render time the client performs one forward pass through \(f_\theta\) to obtain \(\Delta\mathcal{G}(t)\) and updates the canonical splats on-the-fly, incurring negligible CPU overhead and only \(\mathcal{O}(|\mathcal{G}_{\text{base}}|)\) fused-add operations on the GPU.

\subsection{Client–Server Execution Model}
The renderer forms part of a broader \emph{client–server 3DGS pipeline}.  
Training and layer construction occur offline on the server.  
At run-time the server streams the requested LoD layers (and, for animated objects, canonical Gaussian set and MLP weights) to the client.  
In this study we purposely abstract away network latency and throughput, assuming a low-latency, high-bandwidth link so that we can isolate and measure the \textbf{pure rasterization cost} on the client GPU under varying performance constraints.  


\section{Methodology}

Our goal is to investigate the viability and performance behavior of real-time 3D Gaussian Splatting (3DGS) rasterization across a range of GPU classes. As we only have access to a single NVIDIA RTX 4090 GPU, we emulate weaker GPU performance levels via controlled underclocking, using sustained TFLOPS as the primary calibration metric. The following subsections outline our hardware constraints, calibration process, scene setup, and evaluation procedure.

\subsection{GPU Performance Level Emulation via Underclocking}

We emulate four GPU performance levels by restricting the RTX 4090 to operate at four distinct sustained compute performance levels, expressed in TFLOPS (floating-point operations per second). We begin by collecting a list of theoretical FP32 TFLOPS values for a range of modern GPUs from the manufacturer's official specification sheets\cite{nvidia_geforce_specs}. From these theoretical maxima, we estimate the \emph{sustained TFLOPS values} by applying a conservative multiplier of 66.6\%, based on empirical observations in GPU hardware literature \cite{yang_v100_roofline}.

Each target sustained TFLOPS value is then mapped to an underclocked configuration of our RTX 4090. We achieve this using a combination of the following:
\begin{itemize}
    \item \textbf{Power limit reduction}, via \texttt{nvidia-smi -pl}
    \item \textbf{Core clock limit}, via \texttt{nvidia-smi -lgc}
    \item \textbf{Memory clock limit}, via \texttt{nvidia-smi -lmc}
\end{itemize}
These three controls jointly allow us to manipulate the RTX 4090’s performance envelope such that its \emph{measured sustained TFLOPS} aligns with the \emph{estimated sustained TFLOPS} of the target GPU. In addition to matching sustained TFLOPS, we also configure the 4090’s \textbf{core clock}, \textbf{memory clock}, and \textbf{power limit} to approximate the reference GPU’s corresponding specifications. Although these hardware-level values are not matched exactly, we ensure that they fall within the same operating range as the target GPU, while matching the \emph{sustained TFLOPS value}, further increasing the fidelity of our emulation. Sustained TFLOPS are computed empirically by timing floating-point-intensive CUDA workloads (i.e., large General Matrix-to-Matrix Multiplication(GEMM) operations) under each configuration.

Although not an exact replication, this approach implicitly approximates the lower power budgets, memory bandwidths, core frequencies, and compute unit count of the reference GPUs.

The full set of throttling configurations, measurement scripts, and benchmark settings are described in sufficient detail to enable replication of the GPU capability emulation methodology on other hardware platforms.

GPU power usage is logged using nvidia-smi dmon, allowing us to capture time-series power measurements alongside frame rate. From these measurements we derive additional energy-aware metrics including energy per frame and performance-per-watt, which help characterize the efficiency of real-time 3DGS rendering under constrained GPU budgets.

\subsection{Reference GPU Selection and Scope}

To define our four performance levels, we selected four reference GPUs that span a wide spectrum of modern CUDA-compatible hardware, ranging from high-end desktop GPUs to entry-level laptop-class devices. These include the RTX 4090 (flagship), RTX 4070 TI (upper-mid range), RTX 3070 (mid-range) and RTX 3050 (low-range). The goal of this selection is to evaluate how 3D Gaussian Splatting (3DGS) rasterization performance degrades across realistic deployment classes, particularly in edge or client-side rendering scenarios relevant to XR systems.

\begin{table*}[t]
  \centering
  \small
  \caption{Reference devices and the throttling parameters applied to the RTX-4090 emulator.}
  \label{tab:gpu_emulation}
\begin{tabular}{lcccccccccc}
  \toprule
  \makecell[c]{Target GPU} &
  \makecell[c]{Theoretical\\FP32 TFLOPs} &
  \makecell[c]{Estimated\\Sustained\\TFLOPs} &
  \makecell[c]{Nominal\\Power\\(W)} &
  \makecell[c]{Emulated\\Power\\(W)} &
  \makecell[c]{Nominal\\Core Clock\\(MHz)} &
  \makecell[c]{Emulated\\Core \\Clock (MHz)} &
  \makecell[c]{Nominal\\Memory\\Bandwidth\\(GB/s)} &
  \makecell[c]{Required\\Memory\\Clock (MHz)} &
  \makecell[c]{Emulated\\Memory \\Clock (MHz)} &
  \makecell[c]{Measured\\TFLOPs\\after\\Emulation} \\
  \midrule
    RTX\,4090      & 82.58 & 55.05 & 450 & 450 & 2520 & 2520 & 1008 & 10\,501 & 10\,501 (exact) & 53.58 \\
    RTX\,4070 Ti   & 40.09 & 26.73 & 285 & 285 & 2610 & 1125 & 504  & 5\,250  & 5\,001 (–4.7\%) & 26.49 \\
    RTX\,3070      & 20.31 & 13.54 & 220 & 150 & 1725 & 570  & 448  & 4\,667  & 5\,001 (+7.2\%) & 13.49 \\
    RTX\,3050      &  9.10 &  6.07 & 130 & 150 & 1777 & 255  & 224  & 2\,333  & 5\,001 (+114\%) &  6.12 \\
    \bottomrule
  \end{tabular}
\end{table*}

The RTX 4090 serves as our upper-bound baseline, showcasing what 3D Gaussian Splatting can achieve on a flagship desktop GPU.  The RTX 4070 Ti covers the upper-mid segment: powerful yet far more prevalent in consumer systems than the 4090-while the RTX 3070 stands in for the mainstream mid-tier card found in many gaming desktops.  Finally, the RTX 3050 represents the entry-level boundary for real-time 3DGS; although constrained by a 96-bit memory bus and lower SM count, its FP32 throughput and availability in budget rigs make it an informative lower-limit reference.

We deliberately exclude embedded systems such as the NVIDIA Jetson AGX Orin 64GB and standalone VR SoCs (e.g., Qualcomm XR2, Apple M2) from our study. While these platforms are important in mobile and embedded compute, they differ significantly in architecture and software stack. Jetson AGX Orin is primarily designed for edge AI inference and does not support high-performance rasterization pipelines akin to CUDA-based rendering. Standalone VR SoCs like Qualcomm XR2 typically run mobile graphics APIs such as Vulkan or Metal, lack warp-level execution, and are optimized for FP16/INT8 workloads rather than sustained FP32 throughput. Furthermore, existing 3DGS rasterizers such as GSplat are implemented in CUDA, making direct performance comparisons or portability to such platforms non-trivial.

At present, the 3DGS rendering pipeline, particularly implementations based on CUDA such as \emph{gsplat}, appears to be constrained to laptop and desktop class GPUs. Although mobile SoCs such as the Snapdragon XR2 Gen 2 or Apple M2 continue to improve in theoretical throughput, their architectural constraints, thermal envelopes, and lack of CUDA compatibility make them ill-suited for current real-time 3DGS rasterizers without significant reengineering. Our inclusion of the RTX 3050 is thus not meant as a proxy for mobile hardware, but rather to identify the lowest-tier desktop-class GPU on which real-time 3DGS rendering is still achievable using current methods. As such, our study sets a practical lower bound on viable deployment targets for CUDA-based 3DGS pipelines.

\subsection{Scene and Dataset Configuration}

For consistency across tests, we adopt the standard \textit{Garden} scene from the Mip-NeRF dataset family \cite{barron2022mipnerf360}. All evaluations are run at a fixed image resolution of $1920\times$1080. The scene is first trained using our Layered Level-of-Detail (LoD) training method (discussed in the previous section) and subsequently rasterized at multiple LoD settings, each corresponding to a different order of magnitude in the number of Gaussian splats.

\subsection{Gaussian LOD Variants}

To simulate varying scene complexity, we render the same Garden scene using four different numbers of 3D Gaussian splats. These configurations are obtained by adjusting the number of layers in our Layered Level-of-Detail (LOD) optimization, which enables efficient control over the number of active splats. The number of Gaussians used in each LOD tier is as follows:
\begin{itemize}
    \item LOD 0: 580,604 Gaussians
    \item LOD 1: 1,834,311 Gaussians
    \item LOD 2: 2,795,038 Gaussians
    \item LOD 3: 3,448,340 Gaussians
\end{itemize}

For scenes with animations, we use a single animation containing an additional 38,844 animated splats. The total number of splats for each experiment, including animations when applicable, is shown in Table 2.

\subsection{Evaluation Metrics}

For each combination of performance level (4 levels), inclusion of animations, and scene complexity (4 LODs), we measure:
\begin{itemize}
    \item \textbf{Frames per second (FPS)} - primary metric for real-time performance.
    \item \textbf{GPU memory usage} - peak memory allocation during rendering.
    \item \textbf{Power draw, core clock, and memory clock} - logged using \texttt{nvidia-smi dmon} to validate that the hardware operates within the calibrated target envelope.
\end{itemize}

GPU power consumption is recorded using nvidia-smi dmon during each experiment, providing time-series measurements of GPU power alongside frame rate. From these measurements we derive additional energy-efficiency metrics shown in Eqs.~(\ref{eq:energy_per_frame}) and (\ref{eq:perf_per_watt}) that help characterize the relationship between rendering performance and energy usage.

Energy per Frame:

\begin{equation}
E_{\text{frame}} = \frac{P_{\text{avg}}}{\mathrm{FPS}}
\label{eq:energy_per_frame}
\end{equation}

where, $P_{\text{avg}}$  = average GPU power consumption (W) and FPS = frames per second.
This gives energy per rendered frame (J/frame).

Performance per Watt:

\begin{equation}
\eta_{\text{perf}} = \frac{\mathrm{FPS}}{P_{\text{avg}}}
\label{eq:perf_per_watt}
\end{equation}

where, $\eta_{\text{perf}}$ =  represents performance efficiency (frames per second per watt).

All tests are run multiple times, each lasting 2 minutes, to ensure consistency, and the results are averaged to minimize transient noise.

\section{Performance Analysis}
\label{sec:results}

Table~\ref{tab:perf_table} reports the mean and standard–deviation (SD) of the
real–time frame rate obtained for four Gaussian splat counts
(0.6-3.5M splats) across four GPU tiers, both with and without
per–frame animation deformation.\footnote{%
  Splats are reduced by disabling successive LoD layers; the geometric extent
  of the scene is unchanged.}
All tests were rendered at $1920\times1080$ px with identical camera paths.

\begin{figure*}
  \includegraphics[width=0.49\textwidth]{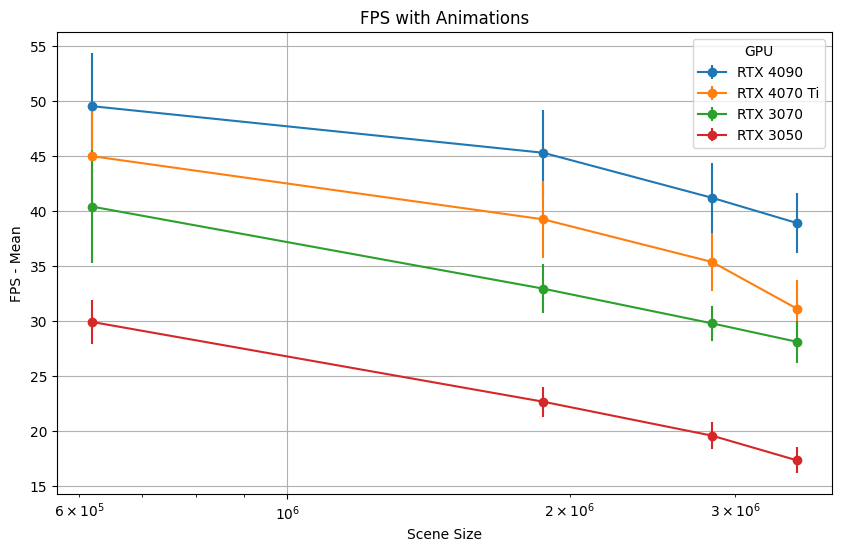}
  \includegraphics[width=0.49\textwidth]{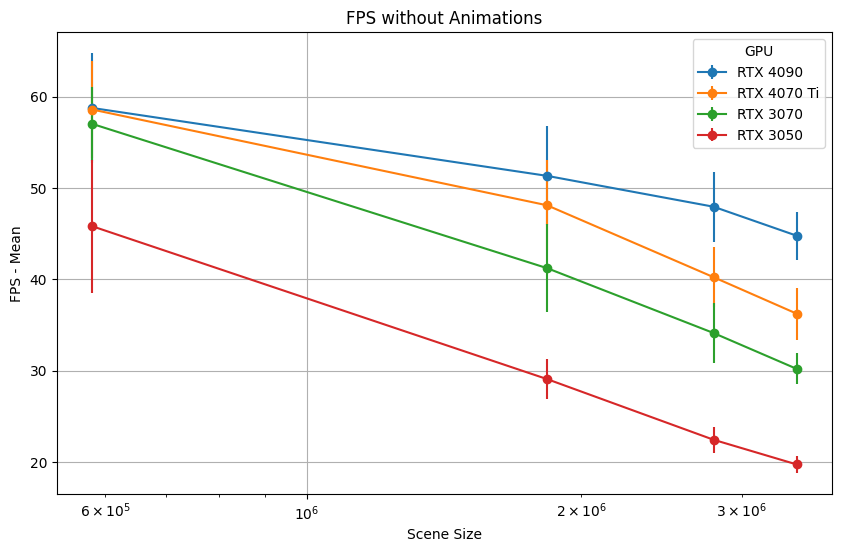}
  \caption{Mean and standard deviation of frame rates across emulated GPU capability tiers and scene sizes, illustrating the performance envelope of real-time 3DGS under constrained compute budgets.}
\end{figure*}

\begin{table}[t]
  \centering
  \footnotesize
  \caption{Measured frame–rate (fps) as a function of GPU tier,
           inclusion of animations, and scene size as number of splats.}
  \label{tab:perf_table}
  \begin{tabular}{llcr@{\,}l}
    \toprule
    \textbf{GPU} & \textbf{Animations?} & \textbf{\# Splats} &
    \multicolumn{2}{c}{\textbf{FPS} (mean $\pm$ SD)}\\
    \midrule
    RTX\,4090 & Yes & 3.49\,M & 38.9 & $\pm$2.7\\
              &     & 2.83\,M & 41.2 & $\pm$3.2\\
              &     & 1.87\,M & 45.3 & $\pm$3.9\\
              &     & 0.62\,M & 49.6 & $\pm$4.8\\
              & No  & 3.45\,M & 44.8 & $\pm$2.6\\
              &     & 2.79\,M & 47.9 & $\pm$3.8\\
              &     & 1.83\,M & 51.3 & $\pm$5.4\\
              &     & 0.58\,M & 58.8 & $\pm$6.0\\[2pt]
    RTX\,4070\,Ti & Yes & 3.49\,M & 31.1 & $\pm$2.6\\
                  &     & 2.83\,M & 35.4 & $\pm$2.7\\
                  &     & 1.87\,M & 39.3 & $\pm$3.5\\
                  &     & 0.62\,M & 45.0 & $\pm$4.8\\
                  & No  & 3.45\,M & 36.2 & $\pm$2.9\\
                  &     & 2.79\,M & 40.2 & $\pm$3.3\\
                  &     & 1.83\,M & 48.1 & $\pm$4.9\\
                  &     & 0.58\,M & 58.6 & $\pm$5.3\\[2pt]
    RTX\,3070 & Yes & 3.49\,M & 28.1 & $\pm$1.9\\
              &     & 2.83\,M & 29.8 & $\pm$1.6\\
              &     & 1.87\,M & 32.9 & $\pm$2.2\\
              &     & 0.62\,M & 40.4 & $\pm$5.2\\
              & No  & 3.45\,M & 30.2 & $\pm$1.7\\
              &     & 2.79\,M & 34.1 & $\pm$3.2\\
              &     & 1.83\,M & 41.2 & $\pm$4.9\\
              &     & 0.58\,M & 57.0 & $\pm$4.1\\[2pt]
    RTX\,3050 & Yes & 3.49\,M & 17.3 & $\pm$1.2\\
              &     & 2.83\,M & 19.6 & $\pm$1.2\\
              &     & 1.87\,M & 22.7 & $\pm$1.4\\
              &     & 0.62\,M & 29.9 & $\pm$2.0\\
              & No  & 3.45\,M & 19.7 & $\pm$1.0\\
              &     & 2.79\,M & 22.4 & $\pm$1.4\\
              &     & 1.83\,M & 29.1 & $\pm$2.2\\
              &     & 0.58\,M & 45.8 & $\pm$7.3\\
    \bottomrule
  \end{tabular}
\end{table}
\subsection{Effect of Animation}
\label{subsec:anim_effect}

Across all GPUs the introduction of time-varying
\emph{deformation‐MLP} splats imposes a clear overhead that grows as
compute capability decreases:

\begin{itemize}
  \item \textbf{RTX~4090.}  Average loss of
        9 fps ($\approx\!15\%$) at \num{0.58} M splats.
  \item \textbf{RTX~4070 Ti.}  $\approx\!13$ FPS drop ($22\%$).
  \item \textbf{RTX~3070.}  $\approx\!17$ FPS drop ($30\%$).
  \item \textbf{RTX~3050.}  $\approx\!16$ FPS drop,
        (\emph{35\%}).
\end{itemize}

\paragraph{Interpretation.}
At every frame the renderer performs an MLP forward pass for the
\(38{,}844\) animated Gaussians, then overwrites their parameters in the
GPU buffer.  The resulting cost combines (i)~extra arithmetic from the
MLP inference and (ii)~additional memory traffic from the per-splat
writes.  Because both costs grow linearly with the number of animated
splats, the impact is most visible on GPUs with fewer cores and narrower
memory buses: the RTX 4070 Ti and 3070 give up a fifth to a quarter of
their baseline frame rate, while the RTX 3050, which is already
compute-constrained, loses over a third.  Tuning the animation budget
(e.g.\ key-frame blend-splats in place of per-frame inference) therefore
could be a strategy for regaining performance on lower-tier hardware, albeit with some more memory cost.





\subsection{LoD Budget vs.\ Scene Scale}

In our experiments LoD reduction is implemented by disabling
fine-detail layers.
Consequently, a 0.6 M–splat Garden scene could correspond to a \SI{2}{\metre}
room if trained with aggressive pruning, or to a full
outdoor capture with coarser parameters.
We chose a splat‐count axis precisely because it isolates the
\emph{rasterizer’s} behavior from the particular capture scale and camera
coverage of any dataset.

\subsection{Performance–Energy Trade-offs}
In addition to frame rate measurements, we analyze the relationship between rendering performance and power consumption. Specifically, we examine FPS–power curves, energy per frame (J/frame), and performance per watt (FPS/W) across the emulated GPU capability tiers. These metrics provide insight into how efficiently real-time 3DGS rasterization utilizes available GPU resources under constrained power budgets.

This analysis highlights how energy-aware metrics can complement traditional graphics performance benchmarks when evaluating real-time rendering systems for edge deployment.

\subsection{Practical Feasibility}

Putting the numbers together:

\begin{itemize}
  \item Real-time (\(\ge\)\,60 FPS) 3DGS is readily achievable on
        4090/4070 Ti/3070 for scenes that fit below
        \(\sim\)\num{600000} splats; even the entry-level
        3050 approaches \SI{46} fps.
  \item At higher LoDs or scene sizes (\(\sim\)\num{3}–\num{4} M splats) the 4090
        remains interactive (\(\sim45\) FPS), whereas a 3070 drops below
        30 FPS and a 3050 gets unusable fps.  Edge-cloud rendering or
        aggressively reduction of LoD is therefore required for lower-end
        hardware at that level of splat density.
Animation overhead is modest on the RTX~4090 (\(<15\%\)) but rises to
roughly \(35\%\) on the RTX~3050.  The slowdown stems from two
per-frame steps: an MLP forward pass for $\sim\!38\,\text{k}$ animated
splats and the subsequent \emph{in-place} update of their parameters in
GPU memory.  Because both steps scale linearly with the number of
animated splats, the impact is amplified on GPUs with fewer cores and
narrower buses.  A practical mitigation, as demonstrated in TC-3DGS \cite{javed2024temporallycompressed3dgaussian}, is to
pre-compute a small set of key-time \textit{blend-splats} and
interpolate them entirely on the GPU at runtime.  This removes the
per-frame MLP pass and replaces scattered writes with a light,
constant-time interpolation, trading a some VRAM (for storing the blendshapes) for a substantial
reduction in arithmetic and memory traffic on mid and low-tier
hardware.
\end{itemize}

Overall, real-time 3DGS is \textbf{already feasible on desktop GPUs down
to RTX 3070}, provided LoD stays under a million visible splats.
Lower-tier GPUs like the RTX 3050 reach usable frame rates only for carefully pruned
scenes or when assisted by a server that transmits coarser LoD layers.
These findings motivate future work on LoD prediction and bandwidth-aware
hierarchies, as well as hybrid client–server rasterization pipelines
that shift fine-detail layers to the edge cloud.

\section{Limitations and Future Work}
\label{sec:limitations}

\subsection{Limitations of the GPU–Performance Emulation}
Table~\ref{tab:gpu_emulation} summarises the clock and power caps used to
throttle our RTX\,4090 so that its \emph{sustained} FP32 throughput
matches that of four reference consumer GPUs.\footnote{%
  We assume \(66\%\) of the vendor-quoted peak TFLOPS based on
  prior measurements that place real workloads at
  60–75\,\% of theoretical peak.}

\paragraph{Bandwidth mismatch.}
The 4090 can only be down-clocked to \SI{405}{MHz}, \SI{810}{MHz}, and \SI{5{,}001}{MHz}. As a result the emulated 3070 and 3050-tiers retain
\(\!\times\!2\)–\(\!\times\!3\) more memory bandwidth than the physical
cards.  For splat counts above \(\sim\!\!1\) M, where production systems
may be bandwidth-bound, our frame-rate numbers for the lower tiers
are therefore optimistic; the extra bandwidth partly compensates for the
compute throttle and inflates FPS relative to real hardware.

\paragraph{SM–count disparity.}
Matching sustained TFLOPS via core–clock reduction does not perfectly reproduce
differences in SM count, L1/L2 cache sizes, or scheduler granularity.
Fewer SMs often translate into poorer warp occupancy and higher register
spills, which our emulation ignores.

\paragraph{Single efficiency point.}
We estimate sustained performance with a fixed 66\%.
Real GPUs vary in terms of FP32 TFLOPs measures compared to their theoretical maxima. Although we chose a conservative estimate of 66\%,
a per-device efficiency sweep would improve the fidelity of sustained GEMM TFLOPs estimation.

\subsection{Study-Design Limitations}
\begin{itemize}
  \item \textbf{Scene diversity.}  We benchmark a single “Garden”–style
        outdoor capture; highly occluded indoor scenes or city-scale
        captures may stress culling and cache differently.
  \item \textbf{CUDA exclusivity.}  Results hold only for
        NVIDIA/CUDA.  Porting the \emph{gsplat} rasterizer to Vulkan or Metal may
        uncover new bottlenecks (e.g., subgroup ballot vs.\ CUDA
        atomics).
  \item \textbf{No network evaluation.}  Our prototype operates in a client server architecture, but all timings are collected with the LoD layers pre-loaded in GPU memory; we do \emph{not} measure latency, packet loss, or bandwidth adaptation.  Consequently, the results reflect pure client-side rasterization cost. 
\end{itemize}

\subsection{Future Work}
\begin{itemize}
  \item \textbf{Bandwidth-aware emulation.}
        Implement software throttles (e.g.\ CUDA \texttt{clamp}) to cap DRAM throughput, matching each
        target GPU’s memory bandwidth more closely.
  \item \textbf{Animation compression.}
        Replace per-frame MLP inference with TC-3DGS style
        blend-splats and time-spline interpolation, reducing both memory
        traffic and MLP inference overhead on mid-tier hardware.
    \item \textbf{Mobile-SoC feasibility at lower LoD.}
        Our data show that an RTX 3050 sustains only 46 fps for
        the 0.5M-splat LoD at 1080p; a standalone VR SoC would
        therefore miss the real-time target at that density.  What
        remains unclear is \emph{how far the splat budget or resolution must be reduced}
        (e.g. 50-100k splats, foveated crops, or quarter-HD
        resolution) before mobile chips become viable, and whether the
        CUDA-centric rasterizer can be expressed efficiently in Vulkan
        compute, or Apple Metal.  A follow-up study could
        prototype the core kernels in Vulkan/Metal, then benchmark modern
        high-bandwidth SoCs (Apple M-series, Snapdragon X Elite) across a
        grid of LoD budgets and resolutions to chart the “mobile viability
        envelope’’ for 3DGS.  
  \item \textbf{Resolution \&\ foveation.}
        Extend benchmarks to 1440 p, 4 K, and foveated rendering to map
        splat density against perceived quality on HMDs.
\end{itemize}

Despite these limitations, our results delineate a clear performance
frontier: \emph{desktop-class GPUs down to RTX 3070 sustain interactive
frame rates for scenes $\le\!1$ M splats, whereas lower-end GPUs like RTX 3050
require aggressive LoD reduction or cloud assistance}. 
Addressing the bandwidth and animation costs identified above is key to
real-world deployment of 3DGS in power-constrained XR clients.

These results also suggest that incorporating energy-efficiency metrics alongside frame rate provides a more complete understanding of the trade-offs involved in deploying real-time 3DGS on energy-constrained devices such as edge or mobile platforms.

\section{Conclusion}

We used a TFLOPs calibration-based emulation study that lets a single
desktop GPU stand in for multiple consumer tiers, then applied it to map
where real-time 3D Gaussian Splatting becomes practical.  Using four
emulated devices—RTX 4090, 4070 Ti, 3070, 3050—and four LoD settings, we
established concrete “viability points’’: the RTX 3070 remains above
60 fps for scenes under roughly one million splats, whereas the RTX 3050
falls short except at the coarsest LoD/scene size.  These data give developers and
system architects a first, device-agnostic chart of which client GPUs
can shoulder 3DGS rasterization unaided and where hybrid or server-side
solutions must step in.

Because the methodology isolates client-side rendering cost while
holding training, networking, and display resolution constant, it can be
re-used to benchmark future LoD schemes, compression methods, and
cross-API ports without rerunning the full capture pipeline.  We hope
the simple process: match sustained TFLOPS, log frame rate per LoD, and
report the break-even scene size, serves as a common benchmarking framework for
upcoming work on making 3DGS truly ubiquitous.
\printbibliography

\end{document}